\documentclass{mn2e}
\usepackage{mnras_cite}
\usepackage{epsfig}
\usepackage{color}
\usepackage{multirow}

\usepackage{graphicx}
\usepackage{subfigure}

\newcommand{\dif}{{\rm d}}
\newcommand{\eq}{\begin{equation}}
\newcommand{\qe}{\end{equation}}
\newcommand{\ar}{\begin{eqnarray}}
\newcommand{\ra}{\end{eqnarray}}
\newcommand{\fig}{\begin{figure}}
\newcommand{\gif}{\end{figure}}
\newcommand{\gtsim}{\hbox{ $\; \buildrel > \over \sim \;$}}

\newcommand{\lcdm}{$\Lambda$CDM}
\newcommand{\Msun}{{\rm M}_\odot}

\begin{document}

\title[Physical constraints on the central mass and baryon content of satellite galaxies.]{Physical constraints on the central mass and baryon content of satellite galaxies.}
\author[M.J.~Stringer, S.~Cole and C.S.~Frenk]{Martin Stringer, Shaun Cole and Carlos Frenk \\ 
Institute of Computational Cosmology, Department of Physics, University of Durham, South Road, Durham DH1 3LE\\
} 
\maketitle

\begin{abstract}
Recent analysis of the Milky Way's satellite galaxies reveals that these objects share a common central mass density, even though their luminosities range over  five orders of magnitude. This observation can be understood in the context of galaxy formation theory by quantifying the factors which restrict the central mass density to a small range. One limit is set by the maximum mass that can collapse into a given region by the hierarchical growth of structure in the standard cold dark matter cosmology.  Another limit comes from the natural thresholds which exist for gas to be able to cool and form a galaxy. The wide range of luminosities in these satellites reflect the effects of supernova feedback on the fraction of cooled baryons which are retained.
\vspace{0.5cm}
\end{abstract}

\section{Introduction}\label{Introduction}

The satellite galaxies of the Milky Way are providing a valuable guide to our understanding of cosmology. Though they constitute only a single statistical sample,  the standard cold dark matter cosmological model (\lcdm) must at least be able to account for their abundance and properties. Attaining such consistency is not proving to be straightforward.

Initial friction between observation and theory concerned the number of observed satellites, which is greatly exceeded by the number of subhalos that \lcdm~predicts in a host halo of the Milky Way's size \cite[and over 1000 citing articles]{Klypin99,Moore99}. A proposed resolution to this was that photoionisation has suppressed the formation of small galaxies in these subhalos \cite{Bullock01,Benson02,Somerville02}.

Observations of satellite galaxies also indicate that these systems boast extremely high mass to light ratios, much greater than are deduced for more luminous galaxies. Mass does not appear to follow light, as might be naively expected. This has been recently quantified by \scite{Strigari08}, leading to a particularly intriguing result.  The central mass of each satellite, as contained within a certain chosen radius (300 pc), appears to be the same for all the satellites, remaining approximately constant over seven orders of magnitude in luminosity.

A suggested explanation for this result, by the authors themselves, is that this common central density may be the hallmark of a common formation time, a suggestion which appeals to the results of N-body simulations. Navarro et al. (1996,1997) found that the central density of halos was proportional to the cosmic background density at the very earliest stages in their assembly,  so halos carry a permanent imprint of their formation epoch.
\pagebreak

Confirmation and extension of these ideas has come from the combination of N-body simulations with semi-analytic models to predict the properties of galaxy populations \cite{Cooper09,LiHelmi09,Maccio09,Okamoto09}. All these calculations identify the early reionisation of gas and minimum temperature for atomic cooling ($10^4$K) as the features which lead to common central mass in satellites.

From N-body simulations of  three different Milky-Way sized host halos, \scite{Kravtsov09} deduces the central masses of subhalos from their accretion epoch and the detailed results of higher-resolution simulations. Central mass is found to be a power-law in subhalo virial mass, $M_{\rm v}$. The empirical result of constant central mass thus corresponds, in turn, to a power-law relationship between virial mass and luminosity.

These connections between this singular observational result and the physics of galaxy formation can be understood with particular clarity by investigating the expected distribution of halos in the central mass -- total mass plane. This distribution is investigated in \S\ref{Distribution}, paying particular attention to the physical explanation behind each feature and generalising the argument to different measurement radii and host halo masses.

Section \ref{Luminosities} takes the argument further to understand how the underlying subhalo masses might map on to the luminosities of the galaxies contained. Rather than demonstrating consistency between these particular results and a particular galaxy formation model, the relative influence of infall, cooling and supernova feedback are explained at a more general level. It is then shown how these processes can lead to the wide range of luminosities observed.

\section{The distribution of satellite halo masses}\label{Distribution}

\subsection{Total halo mass}\label{TotalMass}
The total mass of a satellite halo is defined to be its virial mass when it first enters the host halo. After this point, it is assumed that its central structure is frozen. Any loss of mass would be restricted to the outer shells, not affecting the central mass which is of interest here.  Such changes (in total mass {\em after} accretion) are not relevant because the discussion concerns the temperature of the halo with regard to cooling, which will have mostly taken place while the halo is still isolated. (Substructures are not expected to retain their hot halos long after accretion).

The central mass of a given halo is easily calculated by assuming the standard dark matter density distribution of \scite{Navarro97} which is parameterised in terms of a virial radius, $r_{\rm v}$, and scale radius, $r_{\rm s}$: 
\eq
\frac{M_r }{M_{\rm v}} = \frac{f(r)}{f(r_{\rm v})}\hspace{1cm}\left[f(x)\equiv\ln\left(1+\frac{x}{r_{\rm s}}\right)-\frac{x}{r_{\rm s}+x}\right]\label{M_NFW}
\qe
where $M_r$ is the mass enclosed by radius $r$ and $M_{\rm v} \equiv M_{r_{\rm v}}$. The ratio of the virial radius to the scale radius is referred to as the {\em concentration}, $c\equiv r_{\rm v}/r_{\rm s}$. 

This is a good point to note the definition of virial temperature, $T_{\rm v}$ (in terms of mean particular mass $\mu m_{\rm H}$):
\eq
k_{\rm B}T_{\rm v}\equiv\frac{GM_{\rm v}\mu m_{\rm H}}{2 r_{\rm v}}~.\label{T_v}
\qe
Since this provides a direct relationship between virial radius and mass, halos of a given virial temperature will occupy a clearly defined locus in the $M_r - M_{\rm v}$ plane.
\begin{figure*}
\includegraphics[trim = 8mm 45mm 8mm 45mm, clip, width=\textwidth]{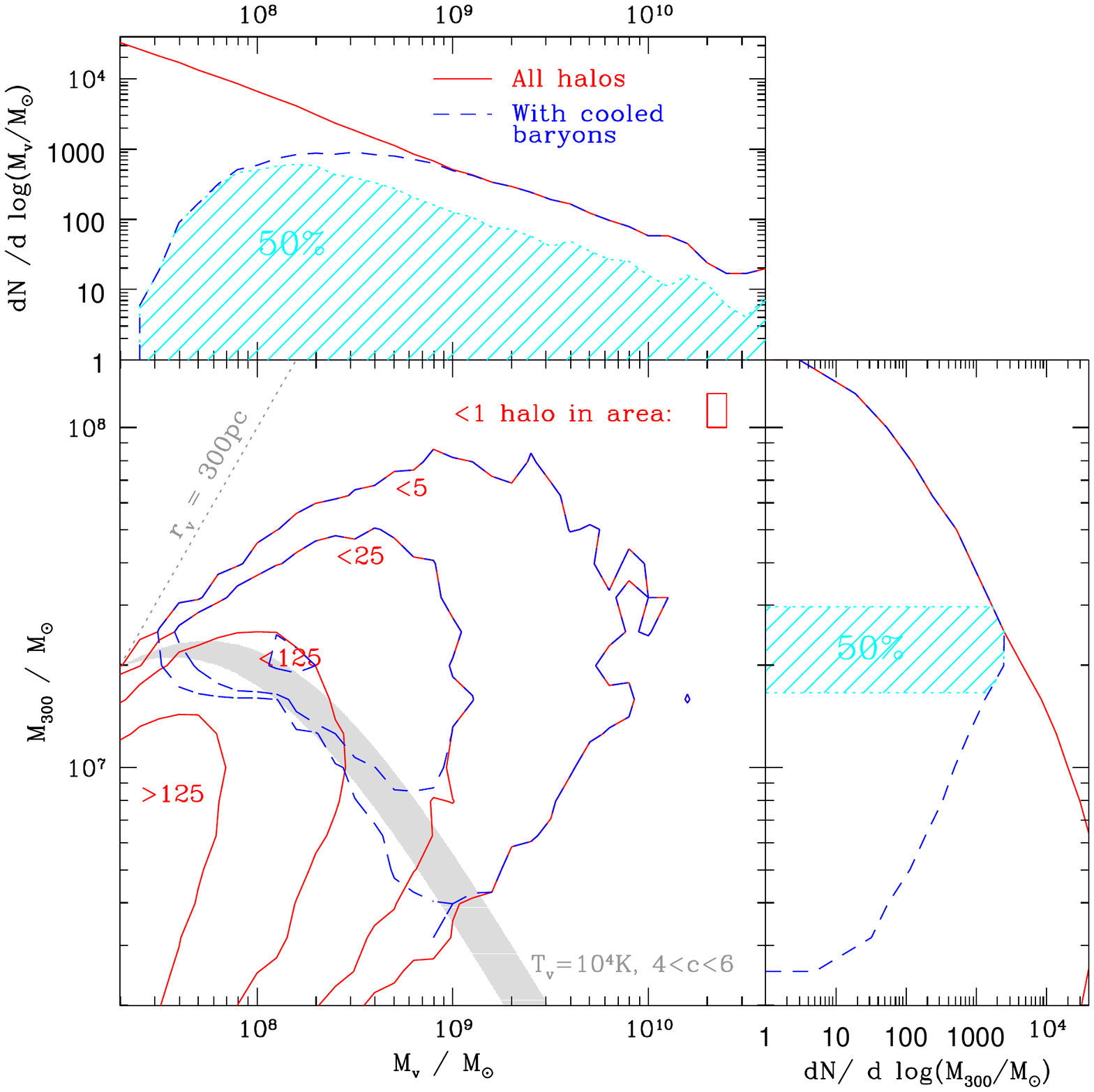}
\caption{The central panel shows theoretical distribution of satellites as a function of their total virial mass, $M_{\rm v}$, at the point of accretion (x-axis) and the mass enclosed within their central 300 pc, $M_{300}$ (y-axis).  Solid (red) lines show all halos. The dashed (blue) lines include only the halos whose temperatures (or their progenitors' temperatures; see Fig.\ref{formation}) have exceeded the threshold for atomic cooling. The locus of halos which are {\em at} this limit is shown as a thick grey line grey, its width allowing for an appropriately small range of halo concentrations, $4<c<6$ (\S\ref{TotalMass}).  The upper panel and right panel show the projection of this distribution onto each axis. The light blue shading in the right panel highlights the central 50\% of the distribution in $M_{300}$ and then, in the top panel, the distribution of these same halos in terms of their virial mass. The mass of the host halo at redshift $z=0$ is $M=2.5\times10^{12}\Msun$. }\label{contours}
\end{figure*}

The subhalo abundance can be predicted, using the Extended Press-Schechter theory \cite{Press74,Bond91}, by building halo merger trees from a conditional mass function \cite{Parkinson08}. At low masses, this mass function tends to a simple power-law:
\eq
\frac{\dif N}{\dif\log(M_{\rm v} / \Msun)} \propto M_{\rm v}^{-\alpha}~,\label{Jenkins}
\qe
as can be seen from the upper panel of Fig. \ref{contours}. This shows the number of satellite subhalos which are predicted to remain within a host halo of mass $2.5\times 10^{12}\Msun$, this lying between two recent estimates\footnote{\scite{Li08} found $2.4\times 10^{12}\Msun$ by comparing the measured recessional velocity \cite{Courteau99} and distance \cite{Stanek98} of M31 to halos in the Millennium simulation \cite{Springel05}. \scite{Guo09} found $2.6\times 10^{12}\Msun$ by using the correlation between stellar mass and halo mass found in both the Millennuim simulation and the SDSS \cite{LiWhite09}, and applying a value of $M_\star=6\times10^{10}\Msun$ \cite{Flynn06}.}  for total mass of the Milky Way \cite{Li08,Guo09}. 

Equation (\ref{M_NFW}) can then be used to find the central mass, $M_r$, at some given measurement radius, $r$. The distribution of halos in the $M_{\rm v} - M_r$ plane can then be visualised, as shown in the main panel of Fig \ref{contours} as a series of (solid red) contour lines.  The projection onto the $M_r$ axis is plotted in the right panel.

\subsection{Understanding the distribution of central masses}

\subsubsection{High mass limit}
The number of halos diminishes rapidly with increasing total virial mass (\ref{Jenkins}), and there is a corresponding natural upper limit for the central mass. However, the exact point at which this occurs on the central mass axis is also dependant on the choice of measurement radius (300 pc in Fig.\ref{contours}).  

A natural cosmological limit might be expected for the quantity of dark matter which can collapse into a region of a given size.  Such a limit can be estimated by considering the characteristic density, $\rho_{\rm c}$, of halos, as defined by \scite{Navarro97} in their proposed density profile:
\eq
\rho(r) = \frac{\rho_{\rm c}}{r/r_{\rm s}\left(1+r/r_{\rm s}\right)^2}~.\label{NFW}
\qe
This characteristic density is found to be proportional to the mean matter density $\bar{\rho}_{\rm M}$ at the very earliest stages\footnote{Specifically, when just 1\% of its mass lies in objects exceeding 10\% of the total final mass.}  in the formation of a halo.
\eq
\rho_{\rm c} \approx 3000\,\bar{\rho}_{\rm M}(a_{\rm form})\hspace{0.5cm}\approx 3000\,a_{\rm form}^3\,\Omega_{\rm M}150\Msun {\rm kpc}^{-3} ~.\label{rho_c}
\qe
Choosing, for example, an early formation time with scale factor, $a_{\rm form}=0.1$, the relationship (\ref{rho_c}) gives a characteristic density of $\rho_{\rm c}=10^8\Msun {\rm kpc}^{-3}$. This can be easily translated into an enclosed mass by considering the integral of (\ref{NFW}) in the limit $r<<r_{\rm s}$.
\eq
M_r \approx 2\pi r_{\rm s}r^2\rho_c \hspace{0.5cm}{\rm or}\hspace{0.5cm}M_{300}\approx \left(\frac{r_{\rm s}}{\rm kpc}\right)10^8\Msun \label{limit}
\qe

It ought now to be no surprise that the distribution of central masses in Fig.\ref{contours} tails off at $M_r\sim10^8\Msun$.  Also, looking at the main panel, it is clear that this high-mass limit in $M_r$ is not constant in $M_{\rm v}$. The reason for this is clear from the dependence on $r_{\rm s}$ in (\ref{limit}): this scale radius will generally increase with $M_{\rm v}$, hence the slanted shape of the outer contour lines at $M_{\rm v}<10^9\Msun$.

Equations (\ref{rho_c}) and (\ref{limit}) show that, to acquire a greater central mass, halos must have formed even earlier than $a_{\rm form}\sim0.1$, or have a much greater scale radius than $r_{\rm s}\sim1$ kpc. The probability of either of these is too low to produce a significant population with higher central mass.

\subsubsection{Low-mass limit - baryonic content}

To connect with the observational findings, we must move on to consider which subhalos may contain luminous baryons.  This can be done by reducing the population to include only those subhalos whose virial temperature has, at some point in their history, exceeded some minimum temperature for cooling.

Such a threshold exists in atomic cooling, at $T\approx10^4$K. Metal line cooling is effective at slightly lower temperatures but this will not be a significant process in metal-poor satellites. Though some early galaxies may form through molecular hydrogen cooling, the process is inefficient \cite{} and we assume that its contribution here is negligible.

The relevance of re-ionisation to the formation of satellite galaxies has been highlighted in the articles listed in \S\ref{Introduction}. After the first stars have formed, background radiation is expected to suppress cooling in halos below a given virial temperature. Simulations indicate that this limit grows with redshift but is about the same order, $10^4$K,  as the atomic cooling threshold \cite{Okamoto08}.

Because the effect of the ionising background is akin to that of the limit imposed by atomic cooling, the arguments in this article will apply equally well to {\em either of these natural thresholds}. It is enough to note that the number of ``galaxies'' in the restricted population must always be interpreted as a {\em maximum} number.

\begin{figure}
\includegraphics[trim = 13mm 55mm 95mm 110mm, clip, width=\columnwidth]{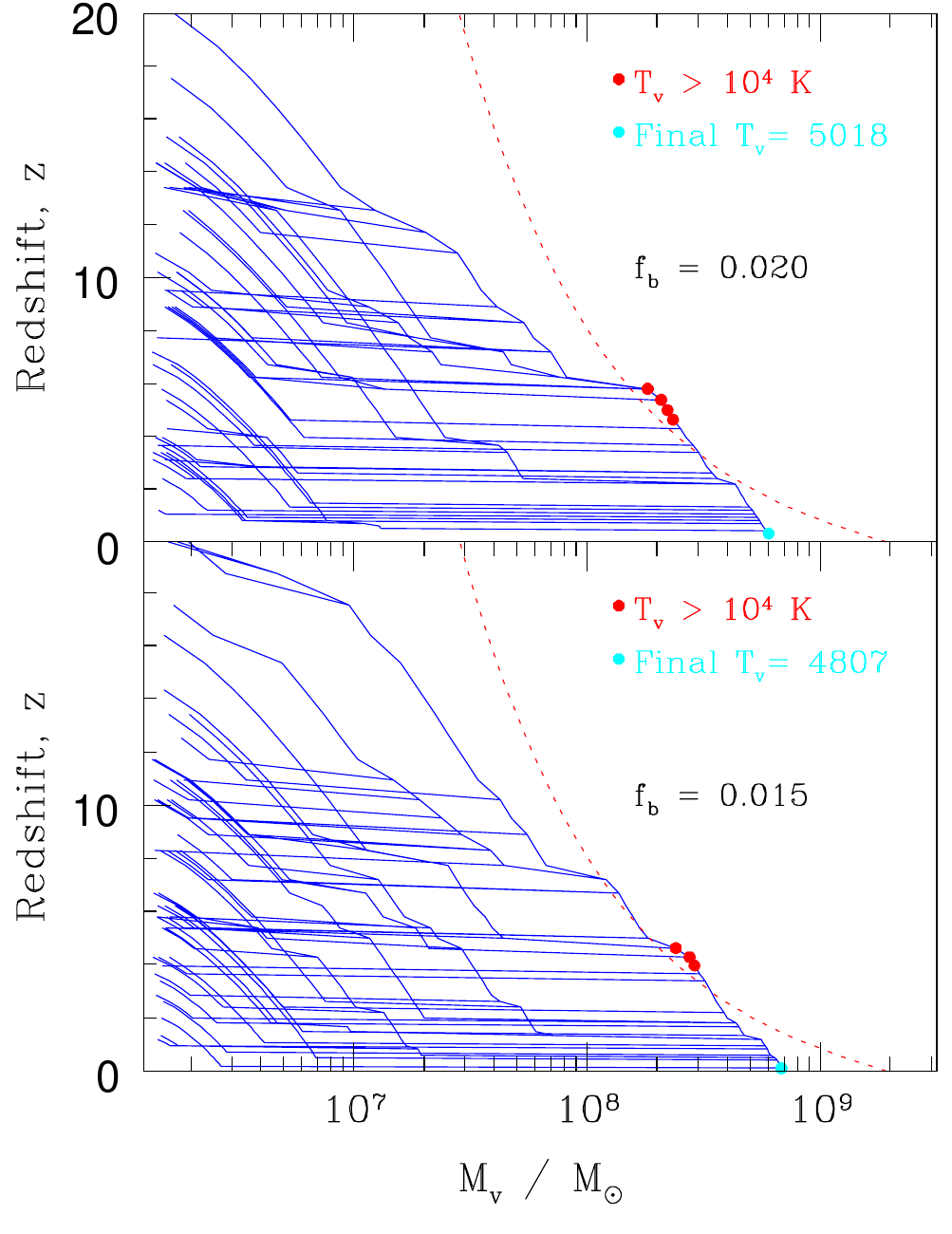}
\caption{The merger tree for two halos in which the final virial temperature is insufficient to allow atomic cooling, yet cooling {\em was} able to occur in their earlier progenitors (red points). Blue lines are drawn between the progenitors of each object and a light blue point highlights the final object. In both these examples, the final cooled baryon fraction, $f_{\rm b}$, is extremely low, due to the short period of time available for the hot baryons to cool by comparison with their free-fall time (\S\ref{InfallTimes})}\label{formation}
\end{figure}

The locus of $T_{\rm v}=10^4$K is shown in Fig. \ref{contours} for an appropriate range of values for the concentration, $c$. It is noticeable that this selected population extends {\em below} the cut-off that is shaded in Fig.\ref{contours}. The reason for this is as follows. Though the halo may have been below the cooling threshold at the time of its accretion, one or more of its progenitor halos exceeded this limit and may have been able to form a galaxy. Two such merger histories from the theoretical subhalo population are shown in Fig.\ref{formation}. 

As pointed out by many previous authors \cite{Kravtsov09,LiHelmi09,Maccio09} the cooling cut-off would prevent a swathe of lower-mass halos from cooling their baryons. Whilst the total halo population (solid lines in the top panel of Fig. \ref{contours}) continues to increase at lower masses, the reduced, cooled baryon population (dashed lines) tails off rapidly at low masses. 

The distribution of central masses for this sub-population is dramatically different. When combined with the natural limit to the {\em maximum} central mass, this leaves a remainder distribution sharply peaked  around a characteristic mass, with half the halos occupying just a tenth of the total logarithmic range (as shown in the right panel of Fig. \ref{contours}).
\begin{figure}
\includegraphics[trim = 62mm 51mm 11mm 96mm, clip, width=\columnwidth]{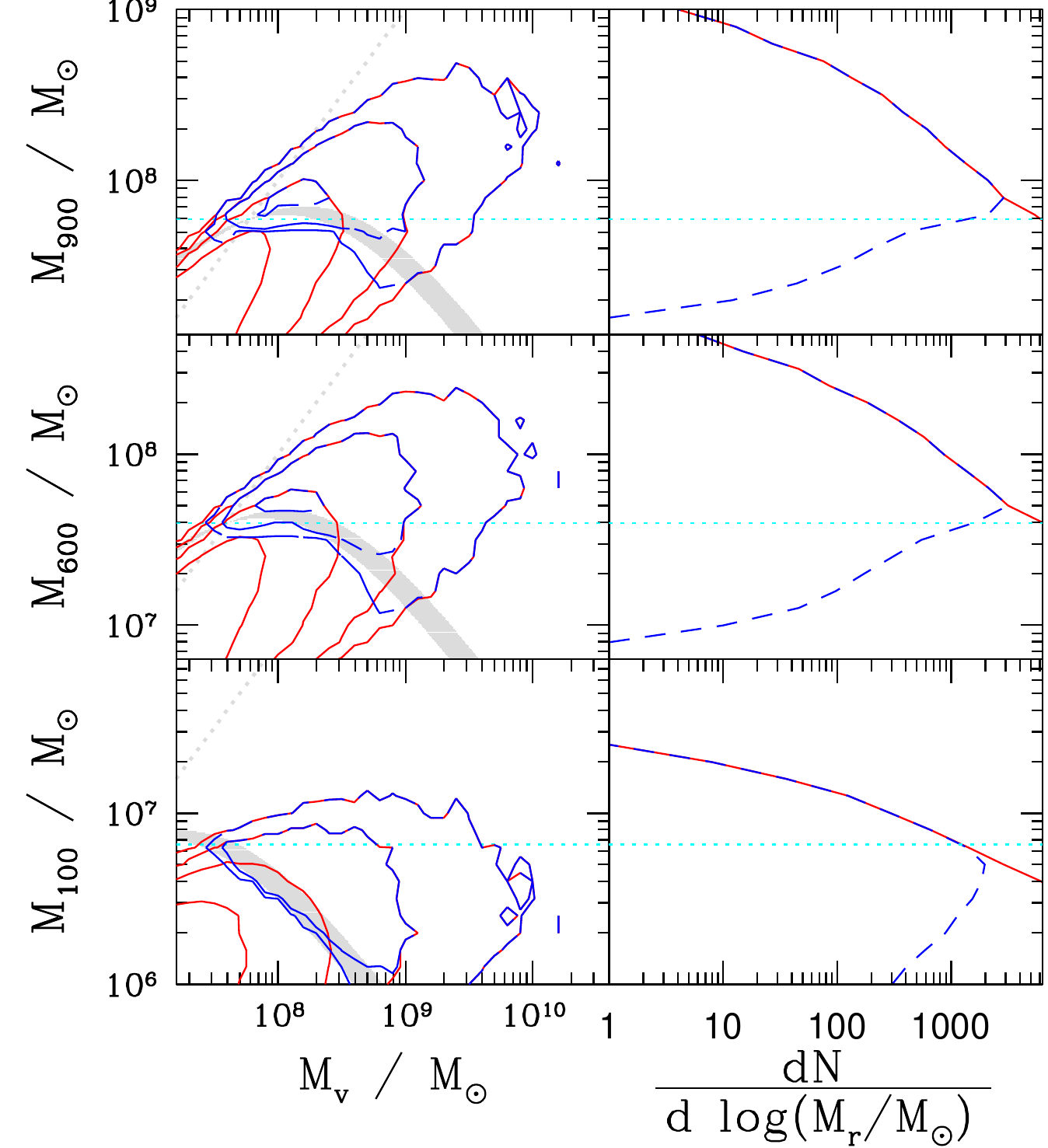}
\caption{The distribution in central mass for three different measurement radii: 100 pc (bottom row) 600 pc (middle) and 900 pc (top). In the panels on the left, the contour lines and key are the same as for Fig.\ref{contours}, with solid red lines showing contours for the entire halo population, and dashed blue lines showing just those halos which can contain cooled baryons. The dotted horizontal line shows the simple estimate (\ref{peak}) for the peak in the central mass distribution. The panels on the right show the projection of this distribution onto the central mass axis. }\label{scale}
\end{figure}

This characteristic mass can be understood graphically. The halos are ``trapped'' between two lines, the dotted line ($r_{\rm v}=300$pc, discussed above) and the curve representing the cooling cut-off. This central mass at which these two lines intersect gives an approximate value for the resulting peak in the distribution:
\eq
M_r^\star \approx \frac{2rk_{\rm B}T_{\rm cut}}{G\mu m_{\rm H}}~. \label{peak}
\qe
Evaluating this for the measurement radius of 300 pc gives the value:
\eq
M_{300}^\star\approx 2\times 10^7\Msun~.
\qe

The physical explanation behind (\ref{peak}) is that the central masses of any satellite population will be tightly contained within two natural limits. At the high-mass end the limit is set by the maximum dark matter mass which can be  enclosed by the chosen measurement radius. At the low-mass end it is set by the requirement that the halo be large enough to have reached a high enough temperature for atomic cooling. This latter requirement allows baryons to fall into the centre and form a galaxy, which may survive the subsequent traumas of accretion into a larger halo and remain as an observable satellite.

\fig
\includegraphics[trim = 62mm 42mm 11mm 138mm, clip, width=\columnwidth]{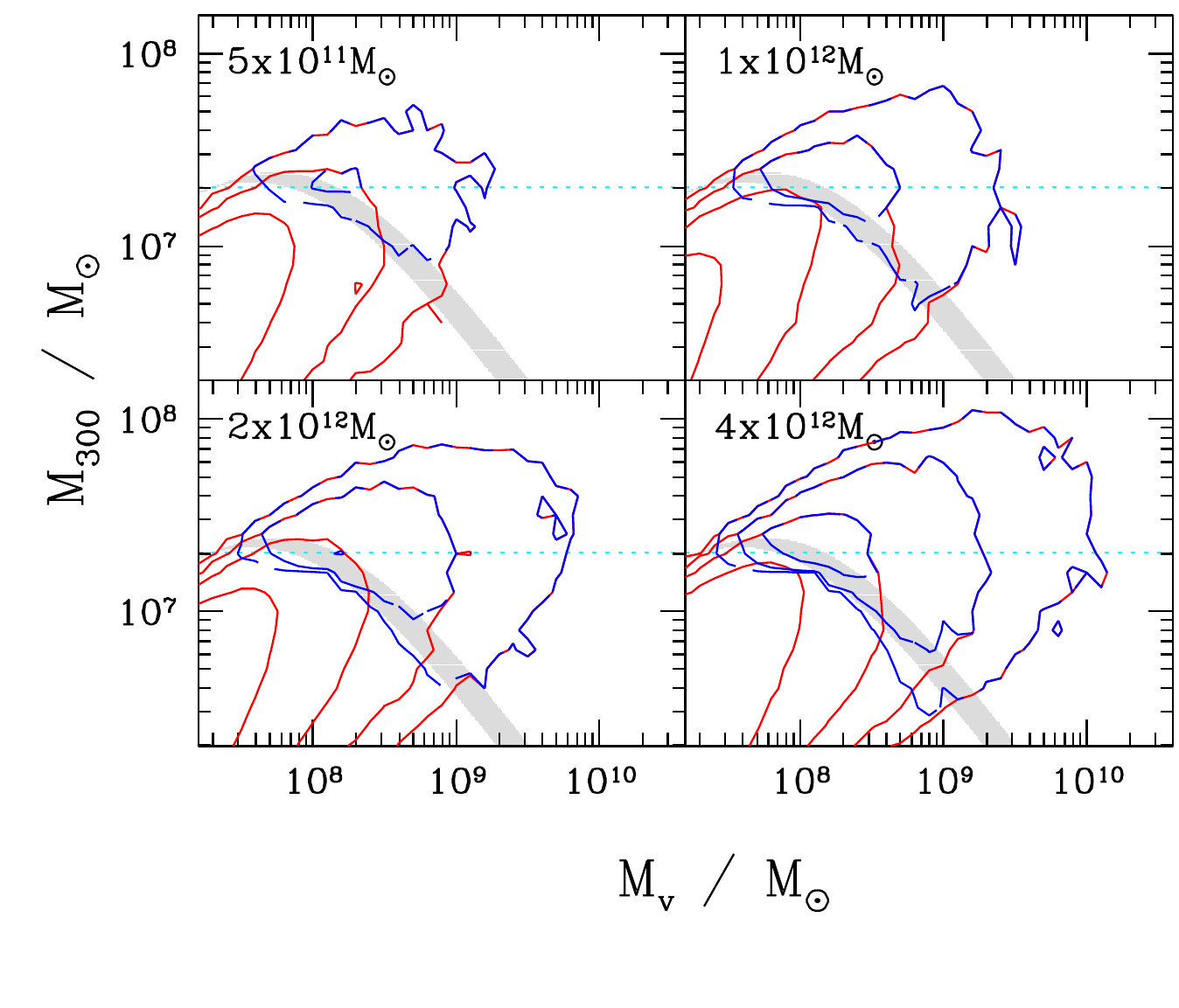}
\includegraphics[trim = 62mm 170mm 11mm 54mm, clip, width=\columnwidth]{massrange.pdf}
\caption{The distribution of satellite halos for a range of host halo masses. 
Each of the top four panels shows a single host mass, with the predicted number of satellites plotted using the same contour lines as Figs. \ref{contours} and \ref{scale}. Red lines show all halos, and blue lines include only those halos whose virial temperature (or their progenitors' temperature) has exceeded $10^4$K. The dotted light blue line shows the characteristic central mass predicted by equation (\ref{peak}).
The lower panel shows the projected distribution on the $M_{300}$ axis for all four host masses and the dotted line again shows the characteristic mass.}\label{massrange}
\gif
     
\subsubsection{The choice of measurement radius}

It is important to widen the discussion and not solely consider the mass enclosed by one particular radius (so far chosen to be 300 pc). Equation (\ref{peak}) is based on the central mass being tightly constrained by the limits of atomic cooling and the maximum mass that can have collapsed into that size of region.  This argument is as valid for a range of choices for enclosure radius as verified by Fig. \ref{scale}.

\subsubsection{Variation of host halo mass}

It is also important to explore the dependance upon host halo mass. Changes in this mass will lead to changes in the distribution, calculated in Figs. \ref{contours} and \ref{scale}, and it is important to check that the explanation given is not exclusive to particular range of halo mass. This is especially interesting because the observational constraints\footnote{\scite{Li08} give $M_{\rm MW}=2.43\times 10^{12}\Msun$ with a 95\% lower confidence limit of $0.8\times 10^{12}\Msun$.} on the mass of the Milky Way's halo are not that tight.

A range of host halo masses is investigated in Fig.\ref{massrange}, which shows that the argument leading to (\ref{peak}) holds equally well for all the sub-halo populations. The peak in the distribution of central mass remains unchanged, the overall population simply increasing with host halo mass as expected.

\section{The expected distribution of luminosities}\label{Luminosities}

\subsection{Timescales for baryonic infall}\label{InfallTimes}
Consider the energy radiation rate {\em per particle}, $\dot{\epsilon}_{\rm r}$ and the thermal energy {\em per particle}, $\epsilon_{\rm th}$, estimated by making standard assumptions:
\eq
\epsilon_{\rm th}=\frac{3}{2}k_{\rm B}T_{\rm v}\hspace{1cm}{\rm and}\hspace{1cm}\dot{\epsilon}_{\rm r}=\Lambda(T_{\rm v},Z)n~.\label{e_th}
\qe
The second equation defines the cooling function, $\Lambda$, which can be found for given temperature $T$ and metallicity, $Z$, by reference to the published tables of \scite{Sutherland93}. $k_{\rm B}$ is Botzmann's constant.  These two quantities provide a cooling timescale for gas at temperature $T$ and gas density $\rho$:
\ar
t_{\rm cool}(n,Z,T) &\equiv& \frac{\epsilon_{\rm th}}{\epsilon_{\rm r}} \\
&=& \frac{3k_{\rm B}T}{2n\Lambda(T, Z)}~,\label{t_cool}
\ra
The value of $\Lambda$ is effectively zero below $T=10^4$K but peaks at temperatures only slightly higher than this. This means that the halo population in which we are particularly interested (those which occupy the region just above the cooling cut-off) will actually have the {\em shortest} cooling times.  To allow reference to Fig. \ref{formation}, we can evaluate (\ref{t_cool}) at the mean density of an isolated halo of mass $M_{\rm v}=3\times 10^8\Msun$ and $T_{\rm v} = 2\times 10^4$K, which gives $t_{\rm cool}\approx 4$Myr.

So, if cooling is so rapid, why do these borderline halos not cool all their baryons into their central galaxies? The time they spend above the cut-off temperature may be short, but it comfortably exceeds the cooling time.  The explanation is that the free-fall time,
\eq
t_{\rm ff}=\sqrt{\frac{1}{G\bar{\rho} } }\hspace{.5cm}\sim 1~{\rm Gyr}~.\label{t_ff}
\qe
greatly exceeds the cooling time in these cases. So, if a halo is fluctuating in temperature as it grows with time, it really has to remain above the threshold for a few Gyr if its central galaxy is to contain a large fraction of the halos total baryonic mass.

\subsection{The importance of outflow}
One process responsible for the ejection of baryons back out of the central galaxy is supernovae explosions. This process can be modelled by the simple expression for the mass of gas which is added to the halo:
\eq
\dot{M}_{\rm out} = \beta\dot{M}_\star~.\label{beta}
\qe
Traditionally \cite{White91}, and still in many recently published models \cite{Bower06,deLucia07}, the ``mass-loading'' factor $\beta$ is estimated for any galaxy on the basis of its maximum rotation speed\footnote{This being a suitable proxy for the escape velocity of the system.}, $v_{\rm disk}$. This is done by assuming that the fraction, $\epsilon_{\rm SN}$, of the available supernova energy which is carried out of the galaxy in outflow is approximately the same for all systems:
\eq
\frac{1}{2}\dot{M}_{\rm out}v_{\rm disk}^2 = \frac{\epsilon_{\rm SN}E_{\rm SN}}{m_{\rm SN}}\dot{M}_\star~.\label{Mout}
\qe
Here, $E_{\rm SN}$ is the total energy available from a supernova and $m_{\rm SN}$ is the mean mass of stars formed per supernova (a consequence of the initial mass function). Equations (\ref{beta}) and (\ref{Mout}) then yield a physically motivated expression for the ratio of outflowing mass to star forming mass:
\eq
\beta = \left(\frac{v_{\rm hot}}{v_{\rm disk}}\right)^2~.\hspace{2cm}\left[v_{\rm hot}^2\equiv\frac{2\epsilon_{\rm SN}E_{\rm SN}}{m_{\rm SN}}\right]
\qe
\subsubsection{Outflow and final baryon fraction}
Neglecting re-cooling for the moment, the total mass ejected at any time is then simply the integral of (\ref{beta}), bearing in mind the hierarchical formation of the final system,  
\ar
M_{\rm out}(t) &=&\int_0^t\sum_i\beta_i(t')\dot{M}_{\star i}(t')\dif t' \label{out1}\\
&=& \left\{\int_0^t  \sum_i\left[\frac{v_{\rm disk}(t)}{ v_i(t')}\right]^2\frac{\dot{M}_{\star i}(t')\dif t'}{M_\star(t) }  \right\} \beta(t)M_\star(t) \nonumber\\
M_{\rm out}&\gtsim&\beta M_\star~.\label{out}
\ra
The subscript, $i$, runs over all the progenitors of the system which exist at time $t'$. The variable $v_i$ is a convenient abbreviation for the value of $v_{\rm disk}$ for progenitor $i$.

The last step to (\ref{out}) used some physical constraints to pin down the value of the integral. If the smaller, progenitor systems have lower rotational speeds than the final galaxy, the ratio of velocities in square brackets will always be greater than one.  However, the contribution from each stage in assembly is weighted by the fractional contribution to the stellar mass, $\dot{M}_\star\dif t/M_\star$, so larger systems (with higher rotation speeds) will make up a greater part of the integral. This will bring the value of the integral close to unity for most formation histories.

\begin{figure*}
\includegraphics[trim = 8mm 50mm 10mm 52mm, clip, width=0.8\textwidth]{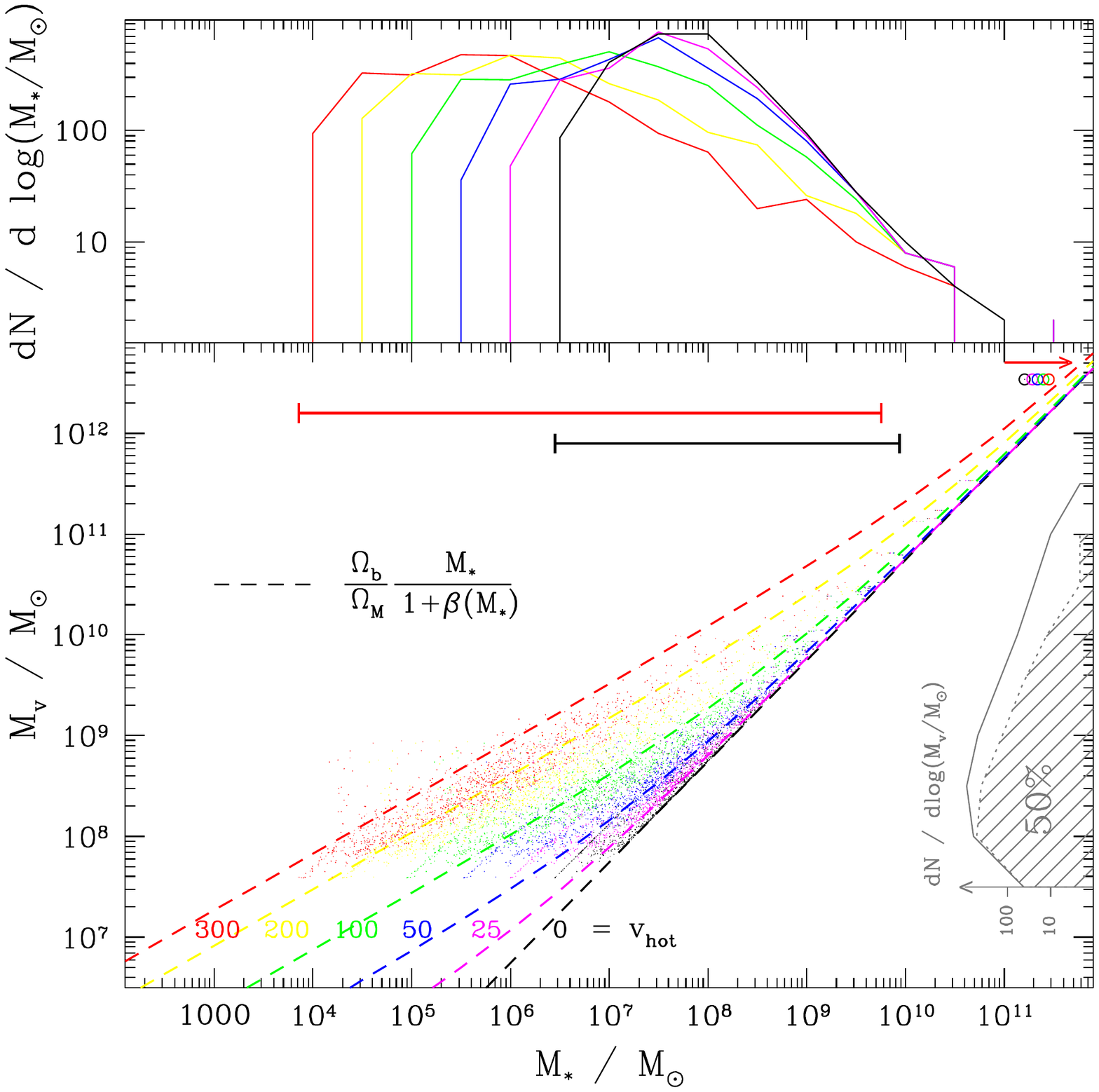}
\caption{The variation in baryonic mass resulting from different strengths of supernova feedback. Dashed lines show the expected analytic relationship (\ref{fb}) between halo mass, $M_{\rm v}$ and the stellar mass contained $M_*$. These different colour show different estimates of supernova efficiency, as encapsulated by the parameter $v_{\rm hot}$. when more mass is ejected. The upper panel shows the corresponding stellar mass function for each mode. Inset in grey on the lower panel, the mass function for the halos is shown again for easy reference. The shaded area shows the halos which occupy the central 50\% of the distribution in $M_{300}$ (see Fig.\ref{contours}). The range that these same halos occupy on the baryonic mass axis is shown as an error bar for the cases of lowest and highest feedback efficiency. The red arrow highlights the {\em increase} in the stellar mass of central galaxy (circled) as feedback increases (opposite to its effect on satellite stellar mass). \vspace{1cm}}\label{frac}
\end{figure*}

Now, in general, the integral (\ref{out1}) gives the total mass ejected over time, {\em not} the total mass which {\em stays} in the hot halo. If the total baryon fraction is close to the cosmological fraction ($\Omega_{\rm b}/\Omega_{\rm M}$), and the mass of gas retained in the galaxy is small compared to the total stellar mass, then $M_{\rm out}+M_\star\approx\left(\Omega_{\rm b}/\Omega_{\rm M}\right)M_{\rm v}$. This leads, with (\ref{out}), to a simple relationship between host halo mass and stellar mass.
\eq
M_{\rm v} \approx \left(\frac{\Omega_{\rm M}}{\Omega_{\rm b}}\right)\frac{M_\star}{1+\beta(M_\star)}\label{fb}
\qe
It is appropriate to write $\beta$ as a function of $M_\star$ because rotational speed is well correlated with stellar mass, both observationally \cite{McGaugh05,Kassin07} and in this model\footnote{Fig.\ref{frac} uses $v_{\rm disk} = (M_\star/10^{10}\Msun)^{0.22}180\,{\rm km\,s}^{-1}$, this being a good match to the rotational speeds of the model systems with which the estimate is being compared}. 

The expression (\ref{fb}) can be used to understand the effect that the outflow process will have when the distribution of satellite {\em halo} masses (Fig.\ref{contours}) is translated into a distribution of the mass of satellite {\em galaxies}. This translation is shown in Fig. \ref{frac} which plots the relationship (\ref{fb}) for a range of efficiencies with which supernova energy is converted to outflow (different values of $v_{\rm hot}$).  

Included in this figure, to provide some basic reassurance as to the validity of (\ref{fb}), are the equivalent estimates derived from an established semi-analytic galaxy formation model \cite{Cole00}, as applied to this very merger tree. This solves a full chain of coupled differential equations, one of which is (\ref{beta}), that interrelate the various evolving properties of a model galaxy population\footnote{The differential equation relating to star formation \cite[eq. 4.4]{Cole00} might be expected to be crucial here, but this mainly controls the cold gas fraction in the galaxies. As galaxies with $M_\star <10^{10}\Msun$ have less than 1\% of their mass in cold gas, this component can be neglected for the purposes of this argument.}. 

This is a good point to redress the first clause of the argument supporting (\ref{fb}), which asks to neglect re-cooling  ``for the moment''. Gas would generally be expected to re-cool onto the galaxy on the order of a free-fall time (\ref{t_ff}). However, in this case of {\em satellite} systems, the hot gas can be stripped off when they are accreted by the host system \cite{Font08}. The question is whether there is usually time, before accretion, for a significant fraction of the ejected gas to re-cool back on to the galaxy.

This question can be addressed using the full galaxy formation model, which includes stripping and recooling. The premise, that the former is the dominant effect ($M_{\rm hot} \approx M_{\rm out}$),  is supported by the full model for lower mass satellites. Larger satellites, on the other hand, can evidently be expected to re-cool {\em some} of their ejected mass before it is stripped. So, the approximation (\ref{fb}) probably underestimates the final stellar mass component for larger satellites, but still provides a useful analytic guide.

With its physical aspects qualified, Fig. \ref{frac} can now be used to understand the effects of feedback on the satellite population. Earlier, Fig. \ref{contours} identified that 50\% of galaxy-containing halos lay within a range of just 0.1 in $\log(M_{300}/\Msun)$. Yet, these same halos were distributed across four orders of magnitude in total virial mass. 

When plotted as a function of their {\em baryonic} mass, these satellites occupy an even greater range due to the effects of feedback. This process disperses galaxies from the same halo population across seven orders of magnitude in baryonic mass for a high (but reasonable) value of outflow efficiency. This provides a plausible explanation for the observed distribution of the masses and luminosities of Milky Way satellites.

\subsubsection{Related implications}

Fig \ref{frac} also emphasises two concepts about the general effects of feedback: 
\begin{itemize}
\item[(i)]{Outflow from supernovae should not be expected to actually reduce the number of satellites, but just reduce the fraction of the total baryons from each halo which are converted to stars. Higher feedback simply shifts the satellite population to lower luminosities, as is clear from the baryon mass functions in the upper panel.  This shift in the satellite stellar mass function {\em reduces} the total number of baryons which are locked up as stars in the substructure.}
\linebreak
\item[(ii)]{Though this ejected gas may be stripped and unable to return to its original galaxy, it {\em can} cool and fall into the centre of its new host halo. This has the interesting consequence that stronger feedback can actually lead to brighter central galaxies, as highlighted in the upper right of the main panel of Fig. \ref{frac}.}
\end{itemize}
\clearpage

\section{Summary}

We have considered the issue of the present-day central density of satellite halos in the context of cold dark matter cosmology theory. By generating satellite populations for a range of host halo masses, the pattern of their distribution in the central mass - total mass plane was understood in terms of physical thresholds. 

The central density must be high enough for the satellite galaxy to survive the effects of reionisation, and/or for the temperature to be high enough to allow atomic cooling. It must also be lower than the natural maximum density which can arise through gravitational collapse from \lcdm~ initial conditions.

These boundaries restrict the population of satellite hosts to a very small range in central mass, but {\em not} in total mass. The explanation for this effect leads to a simple analytic approximation for the central mass (\ref{peak}) given in terms of the general values of host halo mass and enclosure radius. The merger tree calculations support of this conclusion over about an order of magnitude in both factors.

While half the satellites occupy just a tiny portion of the total range in central mass, this same half spans three orders of magnitude in pre-infall virial mass. Further analysis then helps understand how these virial masses might map onto stellar masses, which will explain the range in luminosities observed by \scite{Strigari08}.

By reference to the merger tree formation histories (Fig.\ref{formation}) it can be seen that halos may only be able to cool effectively for {\em part} of their histories. Due to relatively long free-fall times, only a fraction of the baryons that are initially available in these halos will infall into the galaxies. This is one reason why mass-to-light ratios might be high for smaller systems.

A second reason is reheating. In the case of satellites, where hot halo gas can be stripped upon accretion, the traditional argument (that a set fraction of supernova energy is carried out of the galaxy in the reheated gas) is shown to lead to a simple relationship between final stellar mass and halo mass. 

This makes it clear that even quite modest energy conversion will dramatically decrease the fraction of the baryons in satellite halos which can be retained in their galaxies.  For high, but reasonable values of the conversion efficiency, the galaxies which occupy a range in central mass of just  $0.1 \log_{10}(M_r/\Msun)$ are distributed across six orders of magnitude in $M_{\rm baryons}/\Msun$. This is in keeping with the wide range of luminosities observed for the Milky Way's satellites.

\section*{Acknowledgments}

The authors would like to thank Tom Theuns  and Andrew Benson for helpful comments, and the latter for his kind assistance with manipulation of the results from the merger tree calculation. CSF acknowledges a Royal Society Wolfson Research Merit Award.  This work was supported by an STFC rolling grant to the Institute for Computational Cosmology.

\clearpage

\end{document}